\documentclass[preprint]{rmaa}

\usepackage{paralist}

\usepackage{psfrag,color}

\usepackage[latin1]{inputenc}

\title{Epoch-dependent absorption line profile variability in $\lambda$\,Cep} 

\author{J.M. Uuh-Sonda\altaffilmark{1}, G. Rauw\altaffilmark{2}, P. Eenens\altaffilmark{1}, L. Mahy\altaffilmark{2}, M. Palate\altaffilmark{2}, E. Gosset\altaffilmark{2,3}, and C.A. Flores\altaffilmark{1}}

\altaffiltext{1}{Departamento de Astronomia, Universidad de Guanajuato, Mexico.}
\altaffiltext{2}{Institut d'Astrophysique et de G\'eophysique, Universit\'e de Li\`ege, Belgium.}
\altaffiltext{3}{Senior Research Associate F.R.S.-FNRS}

\shortauthor{Uuh-Sonda et al.}
\shorttitle{Line profile variability of $\lambda$\,Cep}

\fulladdresses{
\item Philippe Eenens, Carlos Arturo Flores, Jorge Uuh-Sonda: Departamento de Astronomia, Universidad de Guanajuato, Apartado 144, 36000 Guanajuato, GTO, Mexico.
\item Eric Gosset, Laurent Mahy, Matthieu Palate, Gregor Rauw: Groupe d'Astrophysique des Hautes Energies, Institut d'Astrophysique et de G\'eophysique, Universit\'e de Li\`ege, All\'ee du 6 Ao\^ut, 4000 Li\`ege, Belgium
(gosset, mahy, palate, rauw@astro.ulg.ac.be).}
\listofauthors{J. Uuh-Sonda, G. Rauw, P. Eenens, L. Mahy, M. Palate, E. Gosset, \& C.A. Flores}
\indexauthor{Uuh-Sonda, J.M.}
\indexauthor{Rauw, G.}
\indexauthor{Eenens, P.}
\indexauthor{Mahy, L.}
\indexauthor{Palate, M.}
\indexauthor{Gosset, E.}
\indexauthor{Flores, C.A.}

\abstract{We present the analysis of a multi-epoch spectroscopic monitoring campaign of the O6\,Ief star $\lambda$\,Cep. Previous observations reported the existence of two modes of non-radial pulsations in this star. Our data reveal a much more complex situation. The frequency content of the power spectrum considerably changes from one epoch to the other. We find no stable frequency that can unambiguously be attributed to pulsations. The epoch-dependence of the frequencies and variability patterns are similar to what is seen in the wind emission lines of this and other Oef stars, suggesting that both phenomena likely have the same, currently still unknown, origin.}

\resumen{Reportamos el an\'alisis de una campa\~{n}a de monitoreo espectrosc\'opico multi-\'epoca de la estrella O6\,Ief $\lambda$\,Cep. Observaciones previas reportan la existencia de dos modos de pulsaciones no-radiales en esta estrella. Nuestros datos revelan una situaci\'on considerablemente m\'as compleja. Las frequencias contenidas en el espectro de potencia cambian considerablemente de una \'epoca a otra. No encontramos ninguna frecuencia estable que puedan ser atribuidas inequ\'{\i}vocamente a pulsaciones. La dependencia-temporal de las frequencias y los patrones de variabilidad son similares a los observados en las l\'{\i}neas de emisi\'on del viento en esta y otras estrellas Oef, lo cual sugiere que ambos fen\'{o}menos tienen probablemente el mismo origen, aunque siga todav\'{\i}a sin conocerse.}

\addkeyword{Stars: early-type}
\addkeyword{Stars: oscillations}
\addkeyword{Stars: individual ($\lambda$\,Cep)}
\addkeyword{Stars: variables: general}

\begin{document}
\maketitle

\section{Introduction\label{introduction}}
Line profile variability is a widespread phenomenon among O-type stars, and even more so among O supergiants \citep{FGB}. For single stars, this variability is usually attributed to pulsations, rotational modulations, magnetic fields, structures in the stellar wind, or a mix of these phenomena. The variability of absorption lines, if due to stellar pulsations, can be used to constrain theoretical models of the stellar interior \citep[e.g.][]{Briquet,Godart}. Until recent years, the most robust evidence for pulsations in O-type stars was found in the two O9.5\,V stars $\zeta$\,Oph \citep{Kambe} and HD\,93521 \citep[][ and references therein]{Rauw93521}. Thanks to the high-precision photometry from the {\it CoRoT} mission, $\beta$\,Cep-like pulsations have been found in the O9\,V star HD\,46202 \citep{Briquet}, and probable non-radial pulsations were detected in Plaskett's Star \citep[= HD\,47129, O8\,III/I + O7.5\,V/III,][]{Mahy}. However, these {\it CoRoT} data also revealed a more complex situation as far as earlier main-sequence O-stars are concerned. Indeed, \citet{Blomme} found that the Fourier spectra of the O4\,V((f$^+$)) star HD\,46223, the O5.5\,V((f)) star HD\,46150 and the O8\,V star HD\,46966 are dominated by so-called red noise, i.e.\ stochastic variations whose power increases towards lower frequencies. For these objects, the presence of genuine pulsations could thus not be firmly established. This suggests that the instability that causes the late O-type stars to pulsate does not extend to the earliest spectral types, at least not on the main-sequence.

In this context, the bright O6\,Ief supergiant $\lambda$\,Cep (= HD\,210839) is a very interesting target. To start, it is a bright ($V = 5.08$) and most probably single object \citep{Gies}. $\lambda$\,Cep is one out of a handful of so-called Oef stars in our Galaxy. These objects (including another bright O-star, $\zeta$\,Pup) are characterized by a double-peaked \ion{He}{ii} $\lambda$\,4686 emission line in their spectrum \citep{CL}. The variability of this line in the spectrum of $\lambda$\,Cep has been documented by a number of authors \citep{CF, LC, Grady, McCandliss, Henrichs, Henrichs2, Kaper1, Kaper2}. In the context of the present work, we concentrate mostly on the variability of the absorption lines. Indeed, a previous study of $\lambda$\,Cep's optical absorption line spectrum revealed the likely presence of low-order non-radial pulsations (NRPs) with periods of 12.3 ($l = 3$) and 6.6\,hrs ($l = 5$) \citep{dJ}. However, this result was based on a single line (\ion{He}{i} $\lambda$\,4713) observed during a single five-night multi-site campaign. Since other Oef stars have been found to display a strong epoch-dependence in their variability \citep{bd+60_2522,Oef}, the results of \citet{dJ} needed confirmation. We have thus set up a multi-epoch, multi-site spectroscopic monitoring to re-investigate the absorption line profile variability of $\lambda$\,Cep. 

\section{Observation and data reduction}
Spectroscopic observations of $\lambda$\,Cep were collected during six observing campaigns resulting in a total of 495 spectra. 
Four campaigns (December 2009, June 2010, December 2010 and September 2011) were performed at the Observatoire de Haute Provence (OHP, France) using the Aur\'elie spectrograph fed by the 1.52\,m telescope. The OHP data were taken with a 1200 lines\,mm$^{-1}$ grating  blazed at 5000\,\AA. This set-up covered the wavelength domain from 4460 to 4670\,\AA\ with a resolving power of 20000. The detector was a CCD EEV42-20 with $2048 \times 1024$ pixels of 13.5\,$\mu$m$^2$. Typical exposure times were 10 -- 15 minutes. To achieve the most accurate wavelength calibration, Th-Ar lamp exposures were taken regularly over each observing night (typically once per hour).
The other two campaigns (June 2010 and September 2011) were performed at the Observatorio Astr\'onomico Nacional de San Pedro M\'artir (SPM, Mexico) using the echelle spectrograph mounted on the 2.1\,m telescope. The spectral coverage is about 3800--7300\,\AA\ with a resolving power of 18000 at 5000\,\AA. The detector used for the June 2010 observations  was a Thomson TH7398M CCD with $2048 \times 2048$ pixels of 14\,$\mu$m$^2$, whereas the September 2011 observations were taken with a Marconi 2 CCD with $2048 \times 2048$ pixels of 13.5\,$\mu$m$^2$. Th-Ar lamp exposures were taken about every hour and a half.

The data were reduced using the MIDAS software provided by ESO. The wavelength calibration was done using the Th-Ar lamp exposures. For the Aur\'elie data, the rms error on the wavelength solutions was found to be around 0.0025\,\AA, whilst it is about 0.0080\,\AA\ for the SPM data. A very critical step was the normalization of the spectra. Whilst the Aur\'elie data could be normalized self-consistently over their full wavelength range using a series of continuum windows, this turned out to be much more challenging on the data collected with the echelle spectrograph of SPM. Indeed, because of the strongly peaked blaze of the latter instrument, normalization had to be done on specific orders covering the two lines \ion{He}{i} $\lambda$\,4471 and \ion{He}{ii} $\lambda$\,4542. 

\begin{table*}[h]\centering
  \setlength{\tabnotewidth}{0.5\columnwidth}
  \tablecols{8}
  \caption{Summary of the time sampling of our observing campaigns} \label{journal}
 \begin{tabular}{c c r c c c c c}
    \toprule
    Campaign & Starting and ending dates & $n$ & $<S/N>$ & $\Delta$T & $<\Delta t>$ & $\Delta\nu_{\rm nat}$ & $\nu_{\rm max}$ \\
             & (JD-2450000)              &     &         & (days)    & (days)      & (day$^{-1}$) & (day$^{-1}$) \\
    \midrule
    OHP 12/2009 & 5174.293 -- 5179.385 &  57 & 280 & 5.092 & $1.12 \times 10^{-2}$ & 0.197 & 44.52 \\
    OHP 06/2010 & 5369.406 -- 5375.605 & 100 & 405 & 6.199 & $1.40 \times 10^{-2}$ & 0.161 & 35.60 \\
    SPM 06/2010 & 5370.853 -- 5375.980 &  76 & 250 & 5.127 & $1.31 \times 10^{-2}$ & 0.195 & 38.12 \\
    OHP 12/2010 & 5539.270 -- 5543.433 &  51 & 210 & 4.163 & $1.40 \times 10^{-2}$ & 0.240 & 35.63 \\
    SPM 09/2011 & 5812.788 -- 5820.899 &  95 & 265 & 8.110 & $1.53 \times 10^{-2}$ & 0.123 & 32.68 \\
    OHP 09/2011 & 5825.333 -- 5830.651 & 116 & 340 & 5.318 & $1.40 \times 10^{-2}$ & 0.188 & 35.63 \\
 \bottomrule
  \end{tabular}
\end{table*}

Table\,\ref{journal} provides a summary of our observing campaigns and the characteristics of the sampling. For each campaign, $n$ provides the total number of spectra that were collected, whilst $<S/N>$ indicates the mean signal-to-noise ratio of the spectra. $\Delta$T indicates the total time between the first and last observations, whereas $<\Delta t>$ provides the average time interval between two consecutive exposures of a same night. Finally, Table\,\ref{journal} provides some information in view of the forthcoming Fourier analyses. The last two columns indicate the natural width of the peaks in the periodogram $\Delta\nu_{\rm nat} = (\Delta{\rm T})^{-1}$ as well as $\nu_{\rm max} = (2\,<\Delta t>)^{-1}$ which provides a rough indication of the highest frequency that can possibly be sampled with our time series.

\begin{figure}[!t]\centering
  \includegraphics[width=0.8\columnwidth]{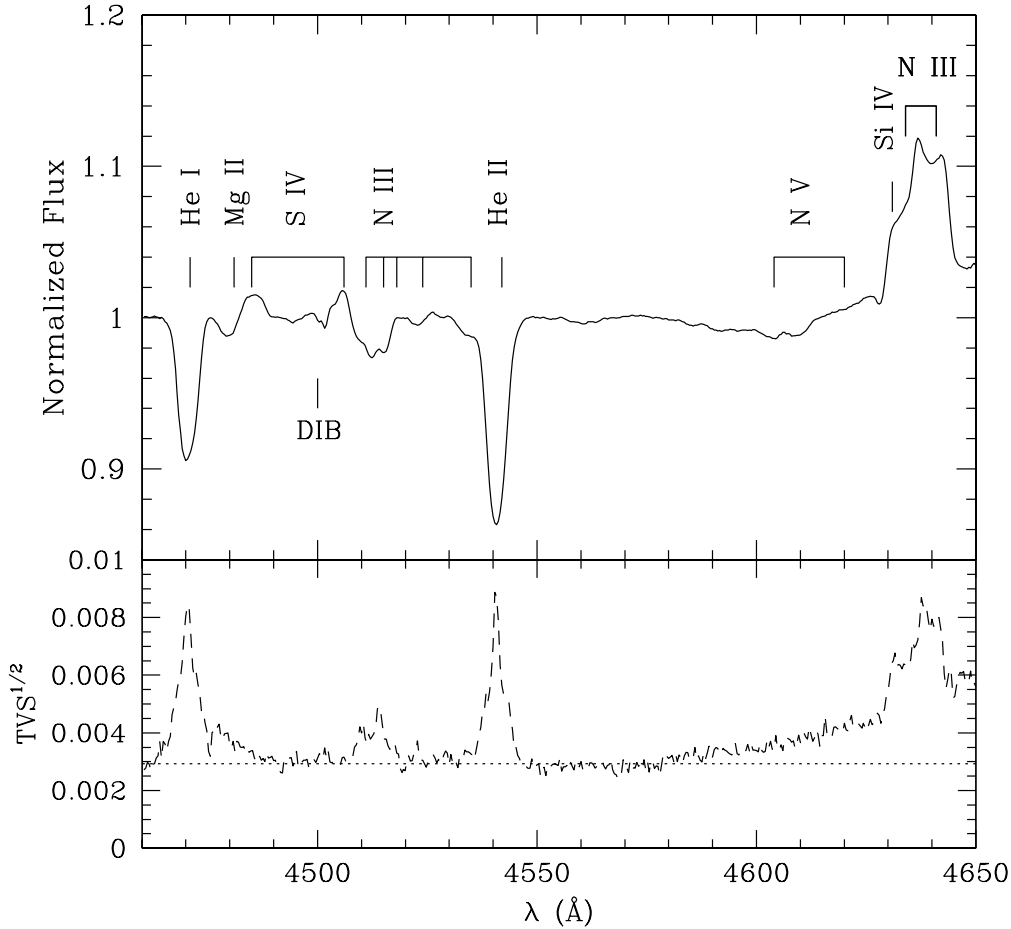}
  \caption{Upper panel: mean spectrum of $\lambda$\,Cep as obtained from the full set of 324 Aur\'elie spectra. The most important spectral features are labelled: stellar lines are identified above the spectrum, whereas the diffuse interstellar band (DIB) near 4500\,\AA\ is indicated below. Lower panel: square root of the temporal variance spectrum of the full set of Aur\'elie spectra. The dotted line yields the 1\% significance level computed according to the mean S/N of the data.}
  \label{OHPallTVS}
\end{figure}

\section{Analysis}
\label{analysis}
To analyze our dataset, we make use of the tools used by \citet{Rauw93521} in their investigation of the variability of HD\,93521. The first step involves computation of the temporal variance spectrum \citep[TVS,][]{FGB} to identify those parts of the spectrum that display significant variability. We then apply a 2D-Fourier analysis on a wavelength-by-wavelength basis over the part of the spectrum where variability is detected \citep[see][]{Rauw93521}. For this purpose, we use the Fourier technique for time series with an uneven sampling designed by \citet{HMM} and amended by \citet{Gosset}. Whenever a significant frequency is found, we prewhiten the data for the corresponding variations, and we then repeat our Fourier analysis on the prewhitened time series. Due to aliasing and the contribution of power at low frequency from a potential red noise component, the choice of the `right' frequency is sometimes non-trivial. This process is performed for up to three frequencies\footnote{As we show below, three frequencies are usually sufficient to account for the power in the Fourier spectrum.}, and allows us to build diagrams showing the amplitude and phase associated to the various frequencies as a function of wavelength. Finally, to evaluate the error bars on the semiamplitudes and phases of the various Fourier components, we have used Monte Carlo simulations assuming that the uncertainties on the normalized flux at a given wavelength and a given time are equal to the noise level of the corresponding spectrum evaluated over a nearby continuum window.

Fig.\,\ref{OHPallTVS} illustrates the mean spectrum and the TVS as computed from the full set of our OHP observations. This figure shows that significant variability is present in the strong absorption lines (\ion{He}{i} $\lambda$\,4471, \ion{He}{ii} $\lambda$\,4542) as well as in the \ion{N}{iii} $\lambda\lambda$\,4634-4642 emission complex. Weaker absorption lines, such as the group of \ion{N}{iii} transitions at $\lambda\lambda$ 4509-4522 are also weakly detected in the TVS, whilst the weak \ion{S}{iv} emission lines are not detected as variable. 

\begin{figure}\centering

  \includegraphics[width=12cm]{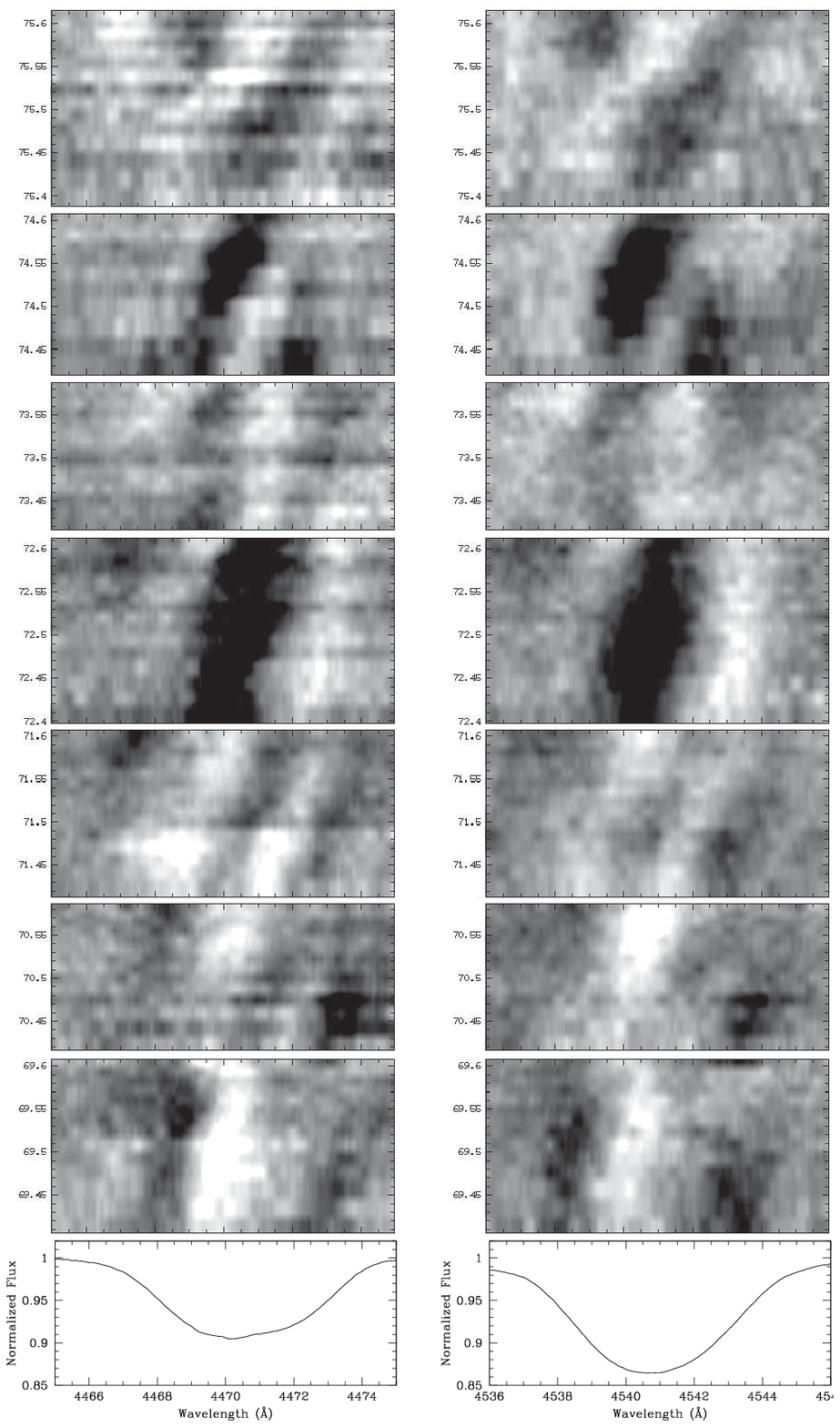}
\caption{Greyscale images illustrating the residuals of individual \ion{He}{i} $\lambda$\,4471 (left) and \ion{He}{ii} $\lambda$\,4542 (right) line profiles with respect to the mean profiles (shown by the lower panels) for the June 2010 OHP data. The labels on the vertical axes of the greyscales yield the date of the observation expressed as HJD-2455300.}
  \label{greyscale}\end{figure}

As an example, Fig.\,\ref{greyscale} illustrates the variability of the \ion{He}{i} $\lambda$\,4471 and \ion{He}{ii} $\lambda$\,4542 lines in the June 2010 OHP data. The variability is clearly seen, as its peak-to-peak amplitude is about 3\%. At first sight, the variations could be consistent with NRPs. Indeed, they mostly consist of red-wards moving substructures similar to what is commonly seen in the line profiles of non-radial pulsators. The variability pattern is not fully constant though, as the widths and strengths of these substructures vary from night to night. Figs.\,\ref{OHPallTVS} and \ref{greyscale} show that, although the variability extends over the full width of the line, the most prominent variations actually occur in the core of the line.

For each subset of our data, we have computed periodograms up to a frequency of 15\,d$^{-1}$ for the \ion{He}{i} $\lambda$\,4471 and \ion{He}{ii} $\lambda$\,4542 lines, respectively over the wavelength domains 4465 -- 4477 and 4535 -- 4546\,\AA. For the OHP data, we have also analyzed the variations of the \ion{N}{iii} $\lambda\lambda$\,4634-42 emission over the 4628 -- 4646 wavelength range\footnote{Due to the strongly peaked blaze of the SPM echelle spectrograph, we were not able to normalize the SPM data of this emission complex with sufficient accuracy.}. This emission triplet is produced in the photosphere, although its strength is dependent on the conditions in the innermost accelerating part of the stellar wind \citep{Rivero}. Therefore the variability of this complex could reflect both the changing conditions in the photosphere (due e.g.\ to pulsations) and at the very base of the wind. The interpretation of its variability is however complicated by the blend with other species such as \ion{Si}{iv} $\lambda$\,4631 and \ion{C}{iii} $\lambda\lambda$\,4647-50.
 
\begin{figure*}[!t]
  \includegraphics[width=0.45\linewidth]{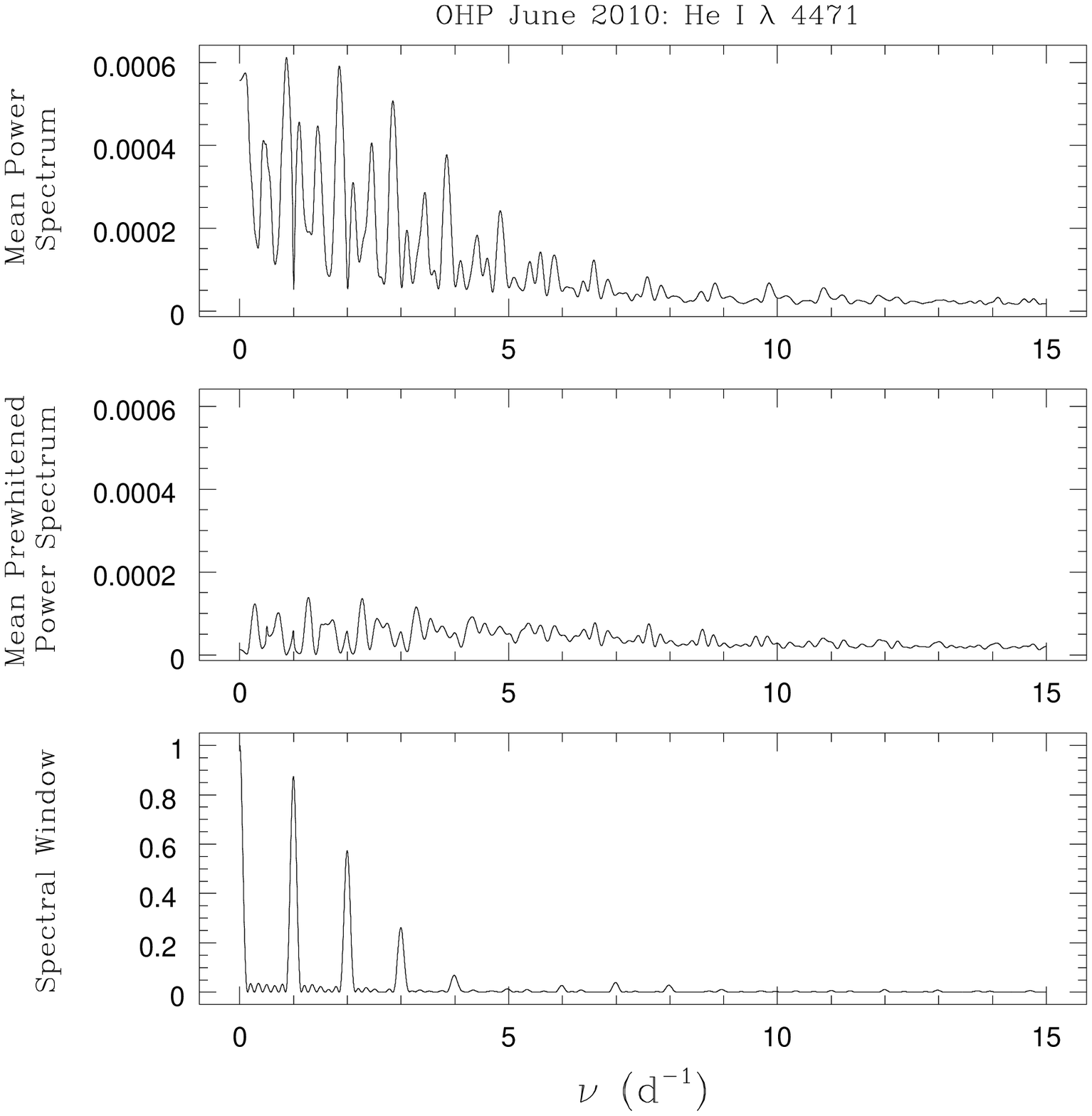}
  \hfill
  \includegraphics[width=0.45\linewidth]{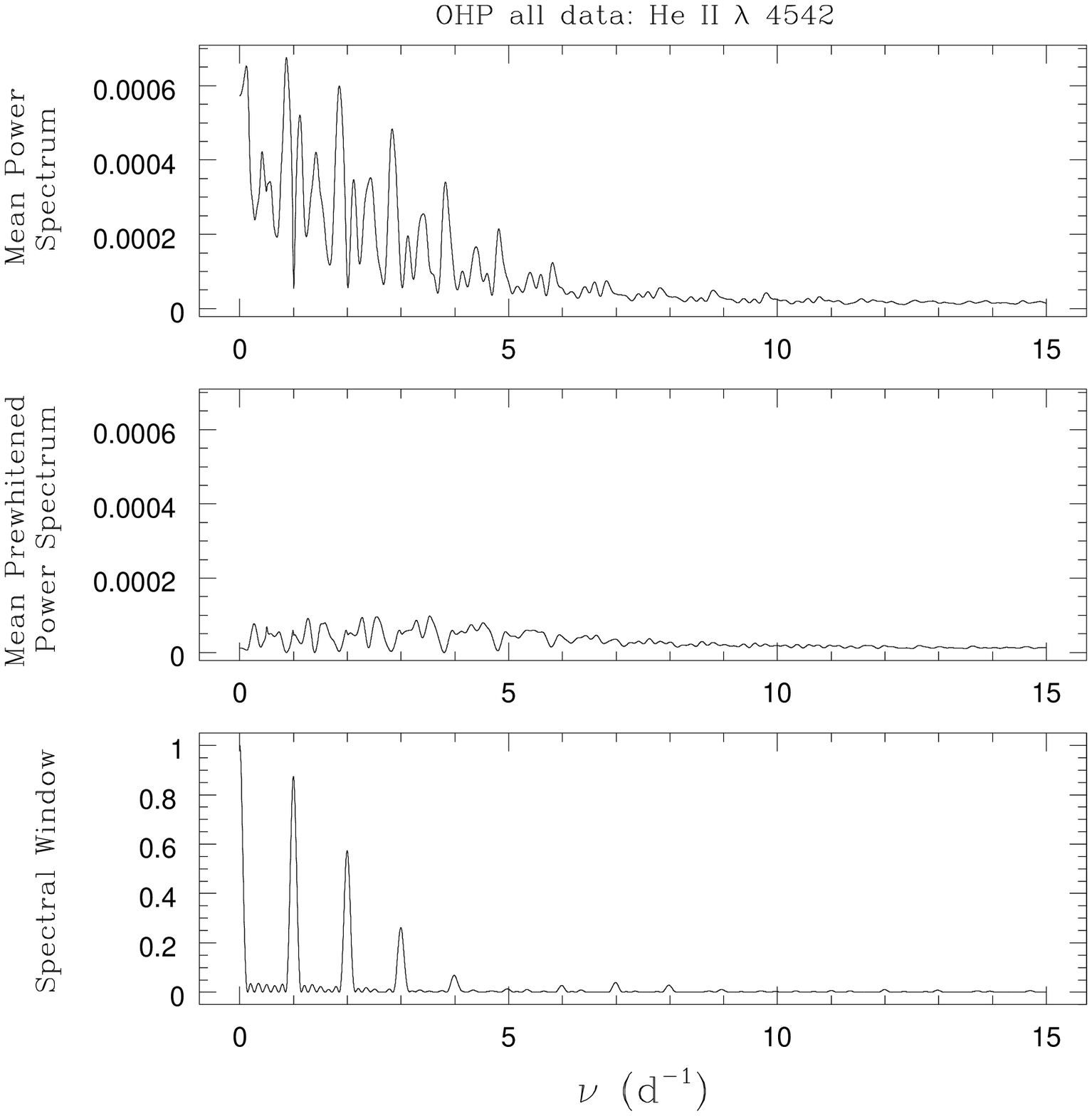}

  \includegraphics[width=0.45\linewidth]{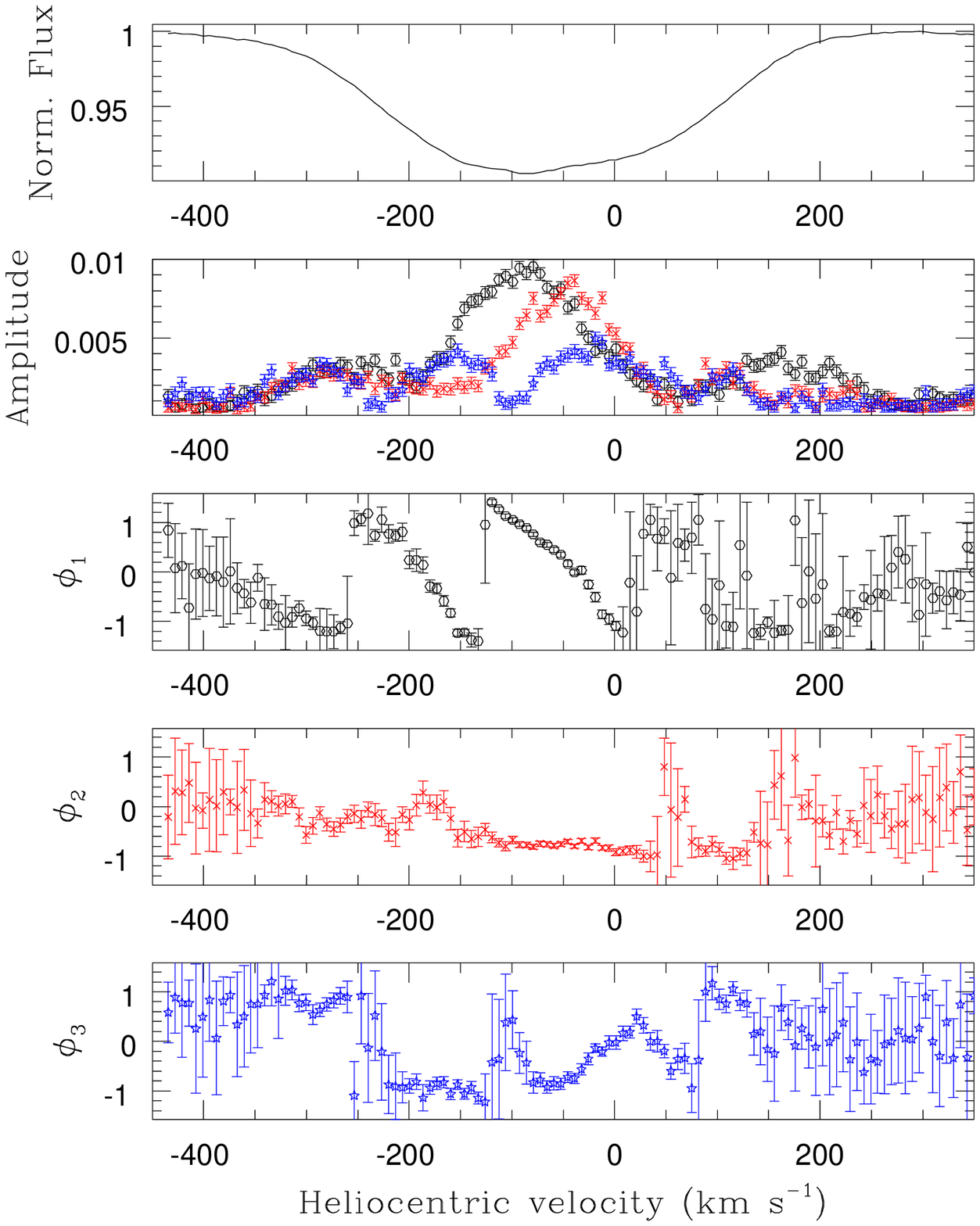}
  \hfill
  \includegraphics[width=0.45\linewidth]{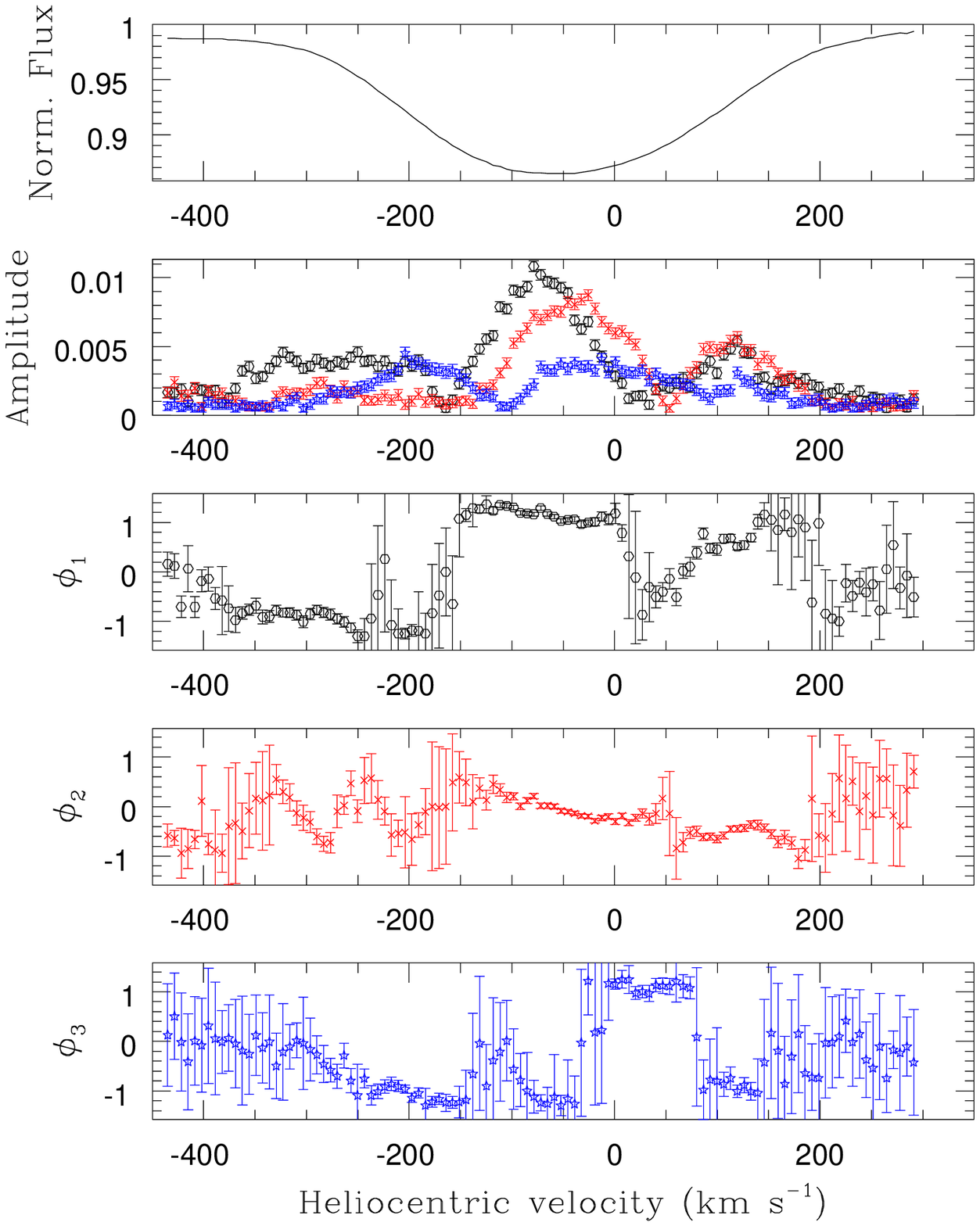}
  \caption{Illustration of the output of our Fourier and prewhitening analysis in the case of the June 2010 OHP data set. The figures on the left correspond to the \ion{He}{i} $\lambda$\,4471 line, whilst those on the right are relative to the \ion{He}{ii} $\lambda$\,4542 line. The top three panels illustrate for each line, the original periodogram, the periodogram after prewhitening of the three frequencies quoted in Table\,\ref{frequencies} and the spectral window (from top to bottom), as a function of frequency. The lower five panels provide the mean normalized line profile, the amplitude of the variations associated with the different frequencies (black hexagone, red crosses and blue stars correspond to  $\nu_1$, $\nu_2$ and $\nu_3$ respectively), as well as their phases modulo $\pi$, as a function of radial velocity across the line profile. The error bars on the amplitude and phase were evaluated via Monte Carlo simulations (see text).\label{June10OHPFourier}}
\end{figure*}

Below we summarize the results obtained in our analyses. Figure\,\ref{June10OHPFourier} illustrates the output of our method for the specific case of the June 2010 OHP data set.

For some of the individual observing campaigns (e.g.\ December 2009, December 2010), the power spectra after prewhitening the dominant frequency are characterized by power at low frequencies which cannot be explained by a simple combination of individual peaks and their aliases. This situation is reminiscent of red noise. In case there is a significant contribution of red noise, it can bias the choice of the right frequency amongst the various aliases. Indeed, the addition of red noise will apparently enhance the strength of the aliases in the lower frequency domain. Moreover, in this case, the prewhitening of a specific frequency overcorrects the effect of this frequency (and overestimates the associated amplitude), as it pumps some power which actually belongs to the red noise.

When all the OHP data are combined, the power spectrum cannot be explained by the combination of a small number of frequencies. Indeed, the actual structures in the power spectrum are broader than expected from the sampling of the data set and prewhitening leaves a substantial power around the prewhitened frequencies that cannot be attributed to their aliases. This result has important implications as far as the stability of the variations is concerned (see Sect.\,\ref{longterm} below). 

Table\,\ref{frequencies} provides an overview of the frequencies that we have detected in our spectroscopic time series of $\lambda$\,Cep. We have checked whether or not our data correctly sample the corresponding cycles. In most cases, the sampling is either very good or even excellent. There are however a few cases where the sampling is poor. These are mostly frequencies near 0.5 or 1\,d$^{-1}$ which are notoriously problematic in single-site ground-based observing campaigns. These frequencies are identified by a colon in Table\,\ref{frequencies}. 

The OHP and SPM campaigns in June 2010 were performed simultaneously, allowing in principle a better coverage of the variations of $\lambda$\,Cep, as the time coverage achieved from both observatories is about 60\% during five consecutive days. Unfortunately, because of the difficulties with the normalization of the SPM data and despite several attempts, it turned out to be impossible to normalize the data from both observatories in a fully consistent way. Therefore, if we combine the time series from OHP and SPM into a single time series, we are dominated by frequencies near 1 and 0.1\,d$^{-1}$, which reflect the artificial variations introduced by the different normalizations. What we can do though, is to compare the frequencies found from the individual observatories. The agreement is quite reasonable: data from both observatories indicate a frequency between 2.12 and 2.24\,d$^{-1}$, as well as a variation with a frequency of 0.87\,d$^{-1}$ or its alias at 2.86\,d$^{-1}$ (either of these frequencies yield good results during the prewhitening process).  

The normalization problems with the SPM data of the September 2011 campaign lead to a periodogram of the \ion{He}{i} $\lambda$\,4471 line dominated by 1\,d$^{-1}$. This is clearly an artefact which impacts our analysis for this spectral line during this specific campaign.

\begin{table*}[h]\centering
  \setlength{\tabnotewidth}{0.7\columnwidth}
  \tablecols{8}
  \caption{Frequencies of the strongest peaks in the power spectra.} \label{frequencies}
 \begin{tabular}{c c c c c c c c c c}
    \toprule
    Campaign & \multicolumn{3}{c}{\ion{He}{i} $\lambda$\,4471} & \multicolumn{3}{c}{\ion{He}{ii} $\lambda$\,4542} & \multicolumn{3}{c}{\ion{N}{iii} $\lambda\lambda$\,4634-42}\\
\cline{2-10}
             & $\nu_1$ & $\nu_2$ & $\nu_3$ & $\nu_1$ & $\nu_2$ & $\nu_3$ & $\nu_1$ & $\nu_2$ & $\nu_3$\\
             & (d$^{-1}$) & (d$^{-1}$) & (d$^{-1}$) & (d$^{-1}$) & (d$^{-1}$) & (d$^{-1}$) & (d$^{-1}$) & (d$^{-1}$) & (d$^{-1}$) \\
    \midrule
   OHP 12/2009  & 2.21 & --   & --   & 2.15 & --   & --   & 2.32 & --   & --  \\
   OHP 06/2010  & 0.87 & 1.41 & 2.12 & 0.87 & 1.40 & 3.80 & 0.56 & 0.85 & 0.30\\
   SPM 06/2010  & 2.22 & 2.86 & 1.60 & 2.24 & 0.63 & 2.90 & --   & --   & --  \\
   OHP 12/2010  & 0.75 & 5.58 & --   & 0.76 & --   & --   & 2.10 & 0.47:& 7.19\\
   SPM 09/2011  & 1.00:& 4.17 & 0.55 & 0.66 & --   & --   & --   & --   & --  \\
   OHP 09/2011  & 0.31 & 3.00 & 0.54:& 0.29 & 1.61 & 3.41 & 0.38 & --   & --  \\

\bottomrule
\tabnotetext{}{A colon behind the value of the frequency means that our time series provides a poor phase coverage of the corresponding cycle.}
  \end{tabular}
\end{table*}

\section{Discussion}
\subsection{The long-term stability of the variability of $\lambda$\,Cep}
\label{longterm}
Although there are a few higher frequencies, the majority of the strongest peaks are located below 3\,d$^{-1}$ (see Table\,\ref{frequencies}). The most obvious candidate for a stable periodicity that emerges from our analysis corresponds to a frequency near 2\,d$^{-1}$ (actually in the range 2.10 -- 2.32\,d$^{-1}$). We note however that, even for this frequency, its visibility in the individual observing campaigns is strongly variable from one epoch to the other. This epoch-dependence can either be genuine, or it can be the result of the sampling of the time series. The latter situation is well illustrated by the June 2010 data. In the SPM time series, the 2.2\,d$^{-1}$ frequency makes up the strongest peak in the periodogram of both helium lines, whilst in the OHP data, it appears only in third position in the power spectrum of the \ion{He}{i} $\lambda$\,4471 line and is missing among the three most important frequencies of the \ion{He}{ii} $\lambda$\,4542 line. 

\citet{dJ} analyzed 169 spectra of $\lambda$\,Cep having a S/N similar to our data, and reported two frequencies, at 1.95 and 3.6\,d$^{-1}$ that they attributed to non-radial pulsations. On the one hand, our data seem to confirm the existence (and long-term stability) of a period near 12\,hours, although the practical problems in correctly sampling this period make it difficult to establish its actual value. Assuming that all detections in Table\,\ref{frequencies} in the range 2.10 -- 2.32\,d$^{-1}$ refer to the same frequency\footnote{If we restrict ourselves to those occurrences that show a behaviour of the phase constant consistent with NRPs (see Table\,\ref{NRPs}), the frequency would be $(2.21 \pm 0.04)$\,d$^{-1}$ for an associated period of $(10.9 \pm 0.2)$\,hr.}, we estimate a value of $(2.19 \pm 0.08)$\,d$^{-1}$, corresponding to a period of $(10.9 \pm 0.4)$\,hr. When considering this frequency as a candidate non-radial pulsation, we note however that the degree $l$ seems to differ from one epoch to the other (see Sect.\,\ref{nonradial}). This casts serious doubt on the stability of the period.  
On the other hand, our data show no evidence of the 3.6\,d$^{-1}$ frequency reported by \citet{dJ}, although such a frequency would be well sampled by our time series. This frequency thus probably reflects a transient phenomenon.   

\citet{Kholtygin} analyzed two series of spectroscopic observations of $\lambda$\,Cep obtained in 1997 (3 nights) and 2007 (4 nights). They found line profile variability at the 2 -- 3\% level in many \ion{He}{i}, \ion{He}{ii} and \ion{H}{i} lines. Unfortunately, only one of the nights of the 1997 campaign was actually suitable for a Fourier analysis which indicated variations of the \ion{He}{ii} $\lambda$\,5412 line with a frequency near 3 -- 4\,d$^{-1}$. The 2007 time series was less densely sampled than the 1997 time series, but \citet{Kholtygin} nevertheless derived a list of 13 different frequencies. A first series of 6 frequencies (0.3, 0.6, 0.7, 0.9, 1.4, 1.6\,d$^{-1}$) was attributed to rotational modulation, whilst the remaining frequencies (2.2, 2.3, 2.5, 2.6, 4.1, 4.5, and 6.9\,d$^{-1}$) were interpreted as non-radial pulsations\footnote{It has to be stressed though that some of these frequencies were apparently only detected over a very narrow wavelength range.}. Many of these frequencies also emerge at least once from our analysis.

\subsection{Non-radial pulsations?}
\label{nonradial}
Establishing the presence of a periodicity in the line profile variability of $\lambda$\,Cep does not necessarily imply that it is due to non-radial pulsations. To check which of the periodicities detected in our analysis are good candidates for NRPs, we have inspected the amplitude and phase diagrams obtained in the course of the prewhitening process. 

Non-radial pulsations are expected to produce a pattern of alternating absorption excesses and deficits that cross the line profile from blue to red as the star rotates. This results in a progressive variation of the phase of the modulation across the line profile. In our data, only a subset of the frequencies identified in Table\,\ref{frequencies} yields a monotonic progression of the phase constant $\phi$ across those parts of the line profile where the amplitude of the variation is large (see e.g.\ the $\nu_1$ frequency of the \ion{He}{i} $\lambda$\,4471 line in Fig.\,\ref{June10OHPFourier}), whilst other frequencies do not show a coherent trend (e.g.\ the $\nu_3$ frequency of the \ion{He}{ii} $\lambda$\,4542 line in Fig.\,\ref{June10OHPFourier}).  

\citet{TS} and \citet{ST} presented linear relations between the observed phase difference between the blue and red line wings and the degree $l$, as well as between the blue-to-red phase difference of the variations in the first harmonics and the absolute value of the azimuthal order $|m|$. These relations were shown to work to first order for spheroidal and toroidal, sectoral and tesseral pulsation modes with $l \leq 15$ and $|m| \geq 2$ \citep{TS}. In our data, we have no detection of the first harmonics, thus preventing us from determining the value of $|m|$. But, we can still attempt to determine $l$ using the relations quoted by \citet{ST}. The results are listed in Table\,\ref{NRPs}. Note that we do not quote values of $l < 2$, as such values are outside the conditions of applicability of the \citet{ST} relations. 
 

\begin{table*}[h]\centering
  \caption{Properties of NRP candidates.} \label{NRPs}
 \begin{tabular}{c l l c c}
    \toprule
$\nu$     & \multicolumn{1}{c}{Line} & \multicolumn{1}{c}{Campaign} & $\frac{|\Delta \phi|}{\pi}$ & $l$ \\
(d$^{-1}$) &      &          &                             &     \\
    \midrule
0.29 & \ion{He}{ii} $\lambda$\,4542 & OHP 09/2011 & $\sim 1.7$ & 2 \\
0.31 & \ion{He}{i} $\lambda$\,4471  & OHP 09/2011 & $\sim 1.7$ & 2 \\
0.54 & \ion{He}{i} $\lambda$\,4471  & OHP 09/2011 & $\sim 2.0$ & 2 -- 3 \\
0.66 & \ion{He}{ii} $\lambda$\,4542 & SPM 09/2011 & $\sim 1.0$ & -- \\ 
0.87 & \ion{He}{i} $\lambda$\,4471  & OHP 06/2010 & $\sim 2.5$ & 3 \\  
1.60 & \ion{He}{i} $\lambda$\,4471  & SPM 06/2010 & $\sim 0.5$ & -- \\
2.15 & \ion{He}{ii} $\lambda$\,4542 & OHP 12/2009 & $\sim 2.5$ & 3 \\
2.21 & \ion{He}{i} $\lambda$\,4471  & OHP 12/2009 & $\sim 2.7$ & 3 \\
2.22 & \ion{He}{i} $\lambda$\,4471  & SPM 06/2010 & $\sim 0.8$ & -- \\
2.24 & \ion{He}{ii} $\lambda$\,4542 & SPM 06/2010 & $\sim 1.0$ & -- \\
2.86 & \ion{He}{i} $\lambda$\,4471  & SPM 06/2010 & $\sim 3.5$ & 4 \\
2.90 & \ion{He}{ii} $\lambda$\,4542 & SPM 06/2010 & $\sim 3.5$ & 4 \\
3.00 & \ion{He}{i} $\lambda$\,4471  & OHP 09/2011 & $\sim 2.5$ & 3 \\
3.41 & \ion{He}{ii} $\lambda$\,4542 & OHP 09/2011 & $\sim 1.5$ & 2 \\
\bottomrule
  \end{tabular}
\end{table*}

The fact that a frequency appears in Table\,\ref{NRPs}, does not necessarily imply that it is truly associated with non-radial pulsations. Indeed, some of the frequencies behave differently for different epochs or lines. For instance, no clear trend is seen in the phase of the 0.63\,d$^{-1}$ frequency in the June 2010 SPM data, although the same frequency shows a trend in the SPM data of September 2011. In a similar way, we note a stepwise variation of the phase constant for the 2.24\,d$^{-1}$ frequency in the \ion{He}{ii} $\lambda$\,4542 SPM June 2010 data, whereas the variation is much more progressive for the 2.22\,d$^{-1}$ frequency in the \ion{He}{i} $\lambda$\,4471 data of the same campaign. In both cases, $|\Delta \phi| \leq \pi$. This behaviour strongly contrasts with the clear monotonic progression of the phase constant over 2.5 -- 2.7\,$\pi$ observed for the 2.15 -- 2.21\,d$^{-1}$ frequency in the OHP December 2009 data. We have checked that this difference in behaviour is robust against the number of frequencies that we prewhiten in a given data set and hence does not arise from interference between the different modes. This result suggests that, after all, we might not be observing the same mode at different epochs. Finally, the phase variation of the 0.87\,d$^{-1}$ frequency in the June 2010 OHP data is different for the \ion{He}{i} $\lambda$\,4471 and \ion{He}{ii} $\lambda$\,4542 lines (monotonic decrease in the former case against roughly constant phase in the latter case). 

It appears thus that there is no evidence for persistent NRPs in $\lambda$\,Cep. Instead, the various modes seem to be transient if they are indeed associated with NRPs.
  
\subsection{Red noise?}
As pointed out in Sect.\,\ref{analysis}, in some cases, prewhitening of a frequency leads to a power spectrum, dominated by low frequencies, but without a distinct individual frequency responsible for the observed power. This situation could indicate that the signal is actually due to a stochastic process, referred to as red noise. \citet{Blomme} reported the detection of such a red noise component in the power spectra of the {\it CoRoT} photometry of early-type O-stars on the main-sequence. These authors suggested that the red noise could come either from a sub-surface convection layer, granulation or wind inhomogeneities. Although speculative, these explanations, especially the first and the last one, could also be relevant for $\lambda$\,Cep. 

\begin{figure*}[!t]
  \includegraphics[width=0.45\linewidth]{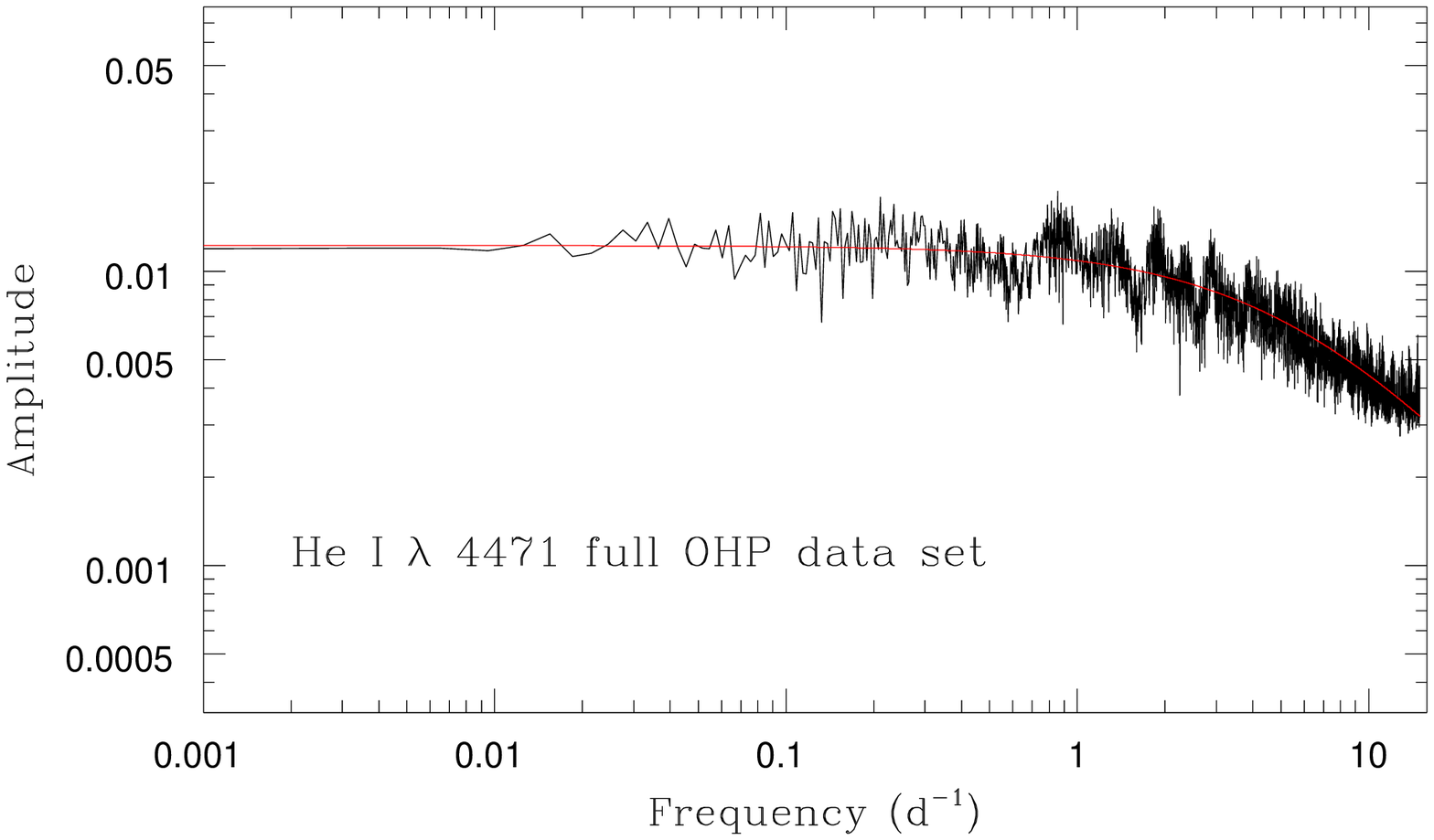}
  \hfill
  \includegraphics[width=0.45\linewidth]{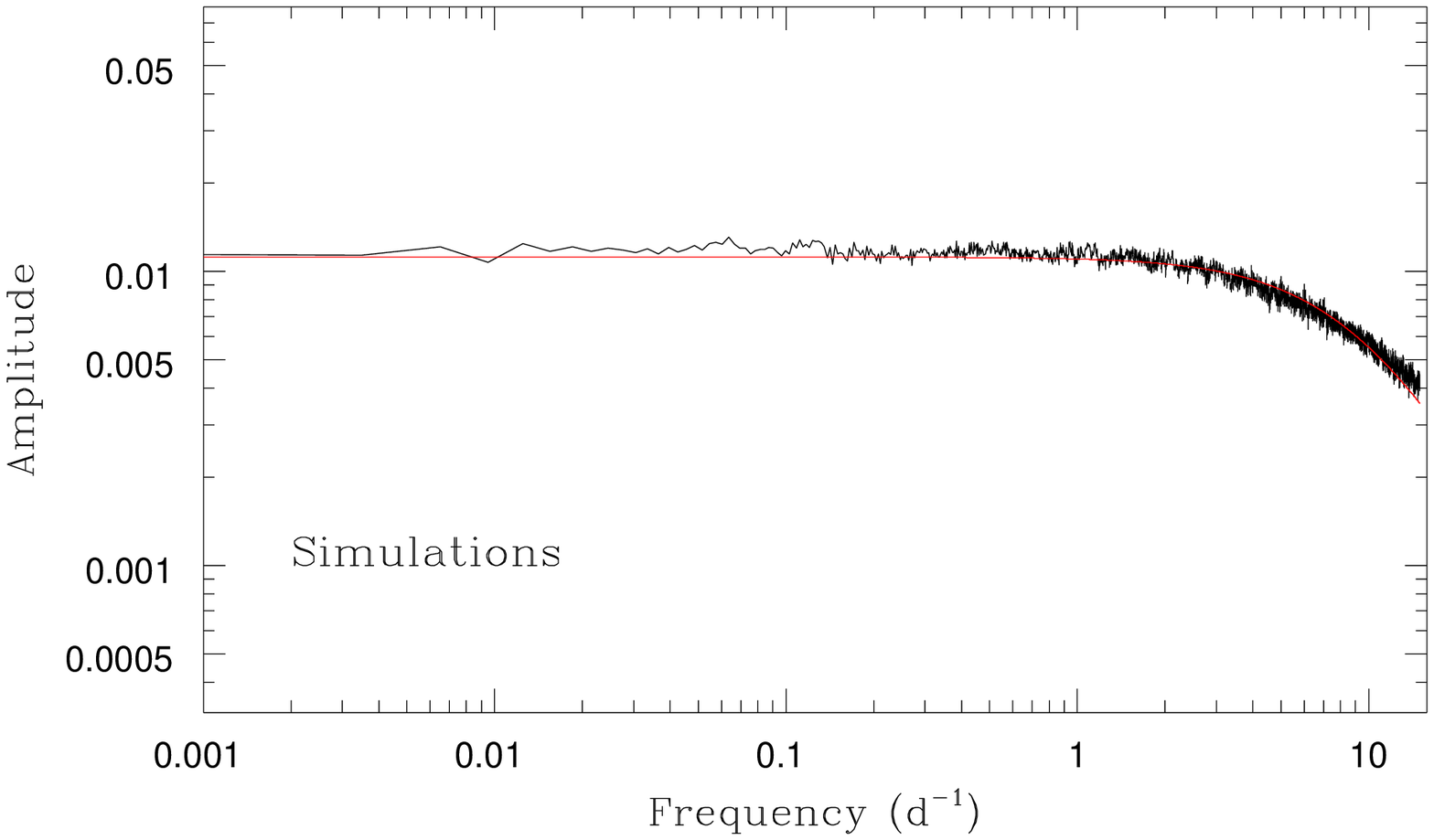}
  \caption{Left: log-log plot of the amplitude spectrum of the full OHP time series of the \ion{He}{i} $\lambda$\,4471 line. The red line shows the best-fit red-noise relation with $A_0 = 0.012$, $\gamma = 1.15$ and $\tau = 0.026$\,day. Right: same, but for simulated data, where all the variations are assumed to arise from red noise. See text for details of the simulations. }
  \label{rednoise}
\end{figure*}

Following, the approach of \citet{Blomme}, which is based on the formalism of \citet{Stanishev}, we have fitted an expression $$A(\nu) = \frac{A_0}{1 + (2\,\pi\,\tau\,\nu)^{\gamma}}$$ to the amplitude spectra of $\lambda$\,Cep. Here $A(\nu)$ is the amplitude derived from the power spectrum, $A_0$ is a scaling factor, $\gamma$ is the slope of the linear part (in a log-log plot) and $\tau$ (in days) is an indication of the mean duration of the dominant structures in the line profile variability. The fits were performed for the amplitude spectra computed between $5 \times 10^{-4}$ and 15\,d$^{-1}$. Above this frequency, the amplitude spectra are mostly dominated by white noise. As an illustration, the left panel of Fig.\,\ref{rednoise} shows a log-log plot of the amplitude spectrum of the \ion{He}{i} $\lambda$\,4471 line in the full OHP time series. 

The best-fit parameters of the red-noise model vary from epoch to epoch and between the various lines. On average, we find $\gamma = 1.22 \pm 0.19$ and $\tau = (0.028 \pm 0.013)$\,day. The $\gamma$ value is intermediate between the values (0.91 -- 0.96) of the three early-type stars investigated by \citet{Blomme} and the value (2.3) of Plaskett's Star studied by \citet{Mahy}. On the other hand, the $\tau$ value is significantly shorter than for any of the above cited stars \citep[0.08 -- 0.17\,day,][]{Blomme,Mahy}. Thus the mean lifetime of the red-noise components is shorter in our case than what is observed in the {\it CoRoT} photometry of O-type stars. 

The left panel of Fig.\,\ref{rednoise} reveals a rather low dispersion of the observed amplitude spectrum about the best-fit $A(\nu)$ relation. At first sight, this is at odds with a red-noise behaviour, where the standard deviation of the periodogram is expected to be equal to the mean of the periodogram \citep[see e.g.][]{TK}, and should thus be significantly larger. The dispersion found here is also much smaller than what was observed in the analysis of the {\it CoRoT} photometry of O-stars \citep{Mahy,Blomme}. To understand the origin of this difference, we need to recall the way our observed amplitude spectrum is obtained. In fact, the observed periodogram is the sum of the periodograms computed at each wavelength bin across the line. Each of the helium lines studied here includes between 110 and 120 wavelength steps. Hence the resulting power spectrum is the mean of 110 -- 120 power spectra. Therefore, if the noise properties of the different wavelength bins are independent, the dispersion of the periodogram is reduced by a factor $\sim \sqrt{110}$ -- $\sqrt{120}$. 

We have generated synthetic light curves dominated by red noise using the recipe of \citet{TK} along with the above formalism for $A(\nu)$ and the average parameters derived above. For the purpose of these simulations, we adopted the same sampling as for the full OHP time series. We further averaged 120 independent power spectra, computed with the same sampling, the same $A(\nu)$ relation, but independent random variables for the Fourier transform of the red noise \citep[see equation (9) of][]{TK}. The latter step was done to mimic the way the observed power spectra are obtained. The resulting amplitude spectrum is shown in the right panel of Fig.\,\ref{rednoise}. As one can see, the simulated amplitude spectrum indeed has a dispersion that is even smaller than what we found in the observed amplitude spectrum. This implies that the intrinsic wavelength width of the red-noise structures in $\lambda$\,Cep is wider than the wavelength step that we have adopted in our analysis. Furthermore, the observed amplitude spectrum displays some excesses that stand out relatively clearly, such as the feature near 2.15 -- 2.21\,d$^{-1}$. The widths of this feature in Fig.\,\ref{rednoise} clearly exceeds the natural width of the entire OHP time series ($1.5 \times 10^{-3}$\,d$^{-1}$), indicating that the feature cannot be due to a single periodicity, but rather arises from a quasi-periodic phenomenon. If we assume, as we have done in Fig.\,\ref{rednoise}, that the bulk of the power is due to red noise, we find that red noise accounts for about half of the observed amplitude of variations at these frequencies. In summary, we conclude that red noise probably contributes to the variations of $\lambda$\,Cep, but additional power could come from a few groups of discrete frequencies. 

\subsection{An alternative scenario?}
The lack of a stable period and/or pattern of variability, as well as the putative presence of a red-noise component suggest a stochastic, chaotic or quasiperiodic phenomenon. This situation likely extends into the variability of the wind lines of stars such as $\lambda$\,Cep. Indeed, \citet{bd+60_2522} and \citet{Oef} found substantial variability in the \ion{He}{ii} $\lambda$\,4686 emission line of other Oef stars with apparent periodicities that exist only over the typical time scale of an observing campaign, but are not stable over the long term. Recently, \citet{HenSud,HenSud2} reached a similar conclusion for the variability of this line in the specific case of $\lambda$\,Cep. They suggested that this situation arises from the presence of numerous, short-lived, localized magnetic loops at the stellar surface, so-called stellar prominences, corotating with the star. These magnetic loops would be the footpoints of wind structures. This explanation could extend to the variability of the photospheric lines studied in this paper if these magnetic loops are associated with spots that rotate across the line of sight, and mimic the line profile variations usually attributed to NRPs. \citet{HenSud2} suggest that the rotation period of $\lambda$\,Cep should be about 4\,d. If the quasi-periodic variations on time scales of 10.9\,hours detected in this paper are indeed due to such spots, one would need about 9 spots on the stellar surface.

To date, there has been no firm detection of a magnetic field in $\lambda$\,Cep: \citet{Kholtygin} inferred an upper limit of 1\,kG on the longitudinal magnetic field. However, if the scenario proposed by \citet{HenSud,HenSud2} is right, the magnetic configuration would be considerably more complex than a dipole, rendering its detection much more challenging \citep[see][]{KochukhovSudnik}.

\section{Conclusion}
The initial goal of our study was to get a deeper insight into the non-radial pulsations of $\lambda$\,Cep. However, our data revealed an unexpectedly complex situation, with no clear evidence for a single set of stable pulsation modes. This situation is very similar to what is observed for the emission lines of this star, as well as of other members of the Oef class, suggesting a common origin. The question as to what causes the observed line profile variability of $\lambda$\,Cep is thus still open. We have investigated the possibility that red noise might be responsible for this phenomenon, but this explanation does not seem to account for the full power. Additional variability likely arises from a large number of transient periodicities that might or might not be due to oscillations of the stellar surface. To make further progress in understanding the variability of this object, more stringent constraints on its magnetic field, regarding both its strength and its geometry, are needed. 

\acknowledgements{This research is supported by a bilateral convention between Conacyt (Mexico) and F.R.S.-FNRS (Belgium), and an ARC grant for Concerted Research Actions, financed by the Federation Wallonia-Brussels. The Li\`ege team further acknowledges further support by the Communaut\'e Fran\c caise de Belgique for the OHP observing campaigns, as well as by Belspo through an XMM/INTEGRAL PRODEX contract.}

\end{document}